\documentclass{svproc}
\usepackage{url}
\usepackage{multirow}
\usepackage{graphicx}

\usepackage{xcolor}

\newcommand{\colora}[1]{\textcolor{violet}{#1}} %magenta
 
\newcommand{\colorc}[1]{\textcolor{teal}{#1}} %blue
\newcommand{\colord}[1]{\textcolor{purple}{#1}} %red
\newcommand{\colore}[1]{\textcolor{brown}{#1}}

\begin{document}
\mainmatter              % start of a contribution

\title{ChatGPT as a tool for User Story Quality Evaluation: Trustworthy Out of the Box?}
\titlerunning{ChatGPT vs Humans}  % abbreviated title (for running head)
\author{Krishna Ronanki \and Beatriz Cabrero-Daniel \and Christian Berger}
\authorrunning{Ronanki et al.} % abbreviated author list (for running head)
\tocauthor{}
\institute{University of Gothenburg, Gothenburg, Sweden\\
\email{krishna.ronanki@gu.se, beatriz.cabrero-daniel@gu.se, christian.berger@gu.se}
}

\maketitle              % typeset the title of the contribution

\begin{abstract}
In Agile software development, user stories play a vital role in capturing and conveying end-user needs, prioritizing features, and facilitating communication and collaboration within development teams. However, automated methods for evaluating user stories require training in NLP tools and can be time-consuming to develop and integrate. This study explores using ChatGPT for user story quality evaluation and compares its performance with an existing benchmark. Our study shows that ChatGPT's evaluation aligns well with human evaluation, and we propose a ``best of three'' strategy to improve its output stability. We also discuss the concept of trustworthiness in AI and its implications for non-experts using ChatGPT's unprocessed outputs. Our research contributes to understanding the reliability and applicability of AI in user story evaluation and offers recommendations for future research.


\end{abstract}

\section{Introduction}

% IMPORTANCE OF GOOD QUALITY user stories IN AGILE SOFTWARE DEVELOPMENT
In agile software development projects, user stories are one of the most widely used notation to express requirements~\cite{lucassen2015forging}. They are considered a very granular representation of requirements that developers use to build new features~\cite{lucassen2016use} as they help to capture \& communicate end-user needs to prioritize \& deliver small, working features in each development cycle~\cite{cohn2004user}. 
 
The quality of user stories is crucial to the success of a development project as they impact the quality of the system design which, in turn, affects the final product~\cite{amna2022systematic}. They provide clear guidance for development efforts, improve communication and collaboration within teams, and help to ensure that development teams have a shared understanding of what needs to be delivered~\cite{9537069}.

% POTENTIAL BENEFITS OF USING AI-BASED TOOLS TO EVALUATE user stories
However, evaluating the quality of user stories manually can be time-consuming. One potential solution for improving Agile software development processes is the integration of automated methods. This can be accomplished through modifications to existing workflows and the implementation of evaluation tools~\cite{humayoun2014user}. Existing methods for automatically evaluating user stories can be relatively fast and efficient, especially when compared to the time required for human evaluation~\cite{Jurisch2017EvaluatingAR}. Natural Language Processing (NLP) has been identified as a potential method for evaluating various aspects of user stories. However, the accuracy and effectiveness of this method can be influenced by factors such as the quality of the training data and the complexity of the user stories under evaluation~\cite{9505355}. Unfortunately, the process of developing and incorporating automated methods for evaluating user stories can be a time-intensive endeavour due to the necessity of training NLP tools to accurately construct algorithms~\cite{Sharir2020TheCO}. 

% INTRODUCE CHATGPT AS A PROMISING AI-BASED TOOL FOR EVALUATING user stories
Developers are increasingly exploring the use of standalone general-purpose applications such as ChatGPT to aid in their software development endeavours. ChatGPT, based on the GPT-3.5 language model, is optimized for dialogue and is capable of answering questions in a human-like text~\cite{zhang2022would}.

% GENERAL BUT WORKS WELL FOR USs
Despite being trained on a large general-purpose corpus and specifically fine-tuned for conversational tasks~\cite{shen2023chatgpt}, it has been observed to perform surprisingly well on specific technical tasks~\cite{choi2023chatgpt}. For this study, we aim at investigating how well a general-purpose language model like ChatGPT performs in evaluating the quality of user stories. 

\section{Background}
% WHAT DO user stories LOOK LIKE
The user story technique is a widely used approach for expressing requirements by utilizing a template that consists of the following elements: ``As a (role), I want (goal), so that (benefit)''~\cite{cohn2004user}. The primary components of a requirement that are captured by user stories are: who is it for, what it expects from the system, and, optionally, why it is important~\cite{cohn2004user}. We follow this user story structure in our study while using the few-shot prompting technique to evaluate the user story quality using ChatGPT. 

Few-shot prompting is a technique where the model is provided with a small number of examples of the task as conditioning in the initial prompt~\cite{radford2019language}. It refers to the ability of language models to learn a new task with limited training samples provided by the user~\cite{perez2021true,liu2023pre}. We used this prompting technique to provide an example to ChatGPT of what a user story should look like structurally before asking it to evaluate the user story on the defined criteria.

% HOW WE MEASURE US QUALITY HERE
The user story quality criteria we used in our study were presented by Lucassen et al.~\cite{aqusa} in their work which focuses on proposing a holistic approach for ensuring the quality of agile requirements expressed as user stories. The approach is comprised of two components: (i) the QUS framework, which is a collection of 13 criteria that can be applied to a set of user stories to assess the quality of individual stories and the set, and (ii) the AQUSA software tool, which utilizes state-of-the-art NLP techniques to automatically detect violations of a selection of the quality criteria in the QUS framework. T\~{o}emets' work investigates whether it is feasible to predict the quality of user stories for monitoring purposes and to determine the correlation between user story quality and other aspects of software development~\cite{Temets2020AnalysingTQ}. The user stories we chose to evaluate as part of our study and the benchmark evaluation scores of the selected user stories using the AQUSA tool were also included in this work.

\section{Method}

\begin{figure}
    \centering
    \includegraphics[width=\linewidth]{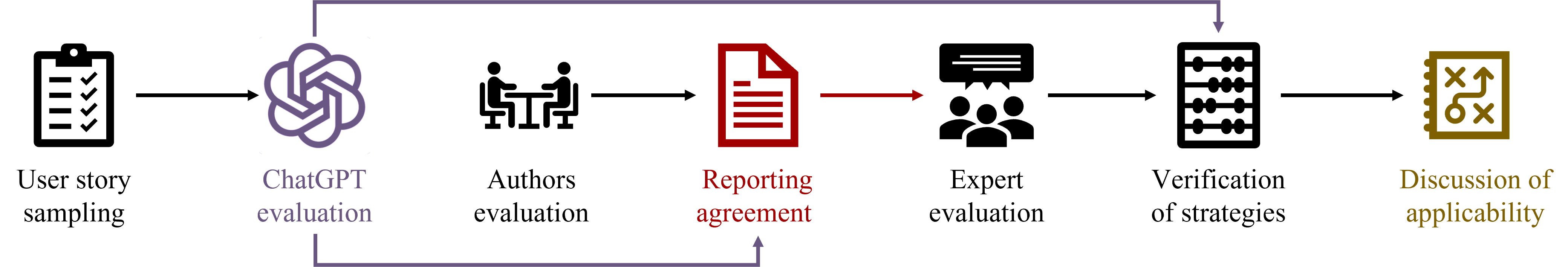}
    \caption{Methodology and verification plan (during workshop)}
    \label{fig:method}
\end{figure}

% GENERAL EXPLANATION
In our study, we performed a comparative analysis of manual and automated evaluation of user stories. Firstly, we assessed the quality of user stories manually, and then we employed ChatGPT for the same task. Our aim was to determine the effectiveness of ChatGPT in evaluating user stories and to compare its performance with human evaluation. The findings of this study are only preliminary, though. We plan to collect further data by involving experienced practitioners present during the XP2023 Research Workshop and repeating the same process with them to also allow comparison between machine-assisted quality assessment and human expertise for quality assessment. This study served as a pilot investigation for the upcoming workshop as reprsented in Figure~\ref{fig:method}.

% HOW WE GOT THE AQUSA AND HUMAN TRUTH
To assess the ability of ChatGPT in replicating human evaluations of user stories, an open-source database presented in T\~{o}emets~\cite{Temets2020AnalysingTQ}\footnote{Visit https://github.com/TanelToemets/Analysing-The-Quality-Of-User-Stories-In-Agile-Software-Projects} was selected as it came with benchmark evaluation of the user stories using the AQUSA tool, which also allows us to refer to an accepted baseline. After retrieving the benchmark, we conducted a double-blinded manual evaluation of the randomly selected set of user stories to assess their quality in terms of atomicity, well-formedness, minimality, conceptual soundness, unambiguity, completeness (full sentence or not), and estimability. The sole criterion for selection was the presence of a benchmark evaluation established using the AQUSA tool. But the evaluation done by AQUSA presented in their study focuses on appraising only the following aspects: atomicity, well-formedness, and minimality.

To evaluate the performance of ChatGPT (March 23 version) for user story quality evaluation, we conducted a series of tests using the one-shot prompting method~\cite{liu2023pre}, a variation of the few-shot prompting method. Specifically, we presented ChatGPT with a set of {criteria, user story} pairs and recorded its responses. To ensure the reliability and consistency of ChatGPT's performance, we repeated this process three times. The evaluation was carried out based on seven criteria presented by Lucassen et al.~\cite{aqusa}. 

% WE CONDUCTED AN EXPERIMENT
Finally, we compared the data from the human evaluation, the AQUSA tool benchmark evaluations and the ChatGPT evaluation and present our findings as tables. The comparison was done for each of the seven criteria and for the overall precision, recall, specificity, and F1 score. The comparison was made to identify any significant differences between the two tools and to ascertain the accuracy of ChatGPT in replicating human evaluation.

% HOW WE GOT THE CHATGPT TRUTH
The results of our experiments raise important issues related to the usability and transparency of ChatGPT's outputs, particularly for non-expert users. In this regard, the discussion section of our paper highlights the need to carefully consider the trustworthiness of ChatGPT's raw outputs and the importance of ensuring that users have the necessary tools to understand and interpret them correctly. By addressing these concerns, we can enhance the usability and effectiveness of ChatGPT as a tool for supporting decision-making in a variety of contexts.

\subsection{Threats to validity}

Validity threats can arise in the benchmark creation process due to the limited scope of evaluation criteria used. The authors of the AQUSA tool evaluated only the atomic, well-formed, and minimal criteria for the sampled user stories. Furthermore, there are concerns about the reliability and accuracy of the evaluation data since it was not provided by the AQUSA authors themselves, but from a master thesis based on Lucasen et al~\cite{aqusa}. Another potential threat to validity is the use of human raters who may not have been experienced practitioners, thus leading to concerns about their reliability. To mitigate this, an independent rating of user stories was conducted, and in case of disagreement, a consensus was reached through a meeting. Moreover, ChatGPT was tested only three times, and further testing could yield different results. Nonetheless, we argue that this is sufficient to assert that developers cannot blindly trust ChatGPT's outputs and integrate them into their Agile software development process. This finding emphasizes the need for cautious and careful consideration when incorporating natural language processing (NLP) tools like ChatGPT into Agile development practices.

\section{Results \& Analysis} \label{sec:results}

\subsection{Comparing the evaluations to the AQUSA benchmark}

% WHAT EVALUATIONS WE HAVE FROM AQUSA
The AQUSA benchmark comprises three key criteria for assessing the quality of a story. The first criterion is whether the story is well-formed, which means it includes a role and the expected functionality, commonly referred to as the means. The second criterion is whether the story is atomic, which implies that it addresses only one feature. The third criterion is whether the story is minimal, which requires that it contains a role, a means, and one or more ends~\cite{aqusa}. 

% AQUSA DOES NOT AGREE WITH OUR EVALUATIONS
Of all pairs \{criteria, user story\}, results showed that human evaluators and AQUSA agreed on only \colord{55\%} of the pairs \{criteria, user story\} as reported in Table~\ref{tab:aqusa}, indicating a moderate level of agreement between the two methods. While human evaluators and AQUSA were in agreement in identifying well-formed and atomic user stories in a majority of cases (81.82\% and 63.64\%, respectively). However, the agreement rate between the two parties dropped significantly when it came to identifying minimal user stories, with only \colorc{18.18\%} agreement observed. The findings of the study indicate that the AQUSA tool exhibits a moderate level of concurrence with human evaluators when it comes to detecting user stories that are well-constructed and atomic in nature, but it currently falls short in identifying minimal user stories.

% WE GOT THE SAME INFO WITH CHATGPT
To enable a fair comparison of results, we conducted evaluations of the same user stories using ChatGPT. The evaluations were performed using two distinct accounts with the history log being cleared between each evaluation. We repeated this process thrice to account for any instability in the results. Table~\ref{tab:aqusa} displays the results of three evaluations conducted to assess the agreement rate and F1 scores of ChatGPT. The findings reveal that ChatGPT demonstrated a consistent agreement rate throughout the evaluations. Furthermore, the F1 scores recorded during the assessments ranged from \colore{81\% to 86\%}. 

\begin{table}[]
\centering
\caption{Percentage of agreement, rounded to 2 decimals, between human evaluations and two tools, AQUSA and ChatGPT, across 11 randomly sampled user stories}
\label{tab:aqusa}
\begin{tabular}{llcccc}
\hline
& & \textbf{AQUSA} & \multicolumn{3}{c}{\textbf{ChatGPT}} \\
\textbf{Type} & \textbf{Metric} & Benchmark & Run 1 & Run 2 & Run 3 \\
 \hline
\multirow{3}{*}{Criteria} & Well-formed & 81.82\% & \colora{81.82\%} & \colora{81.82\%} & \colora{81.82\%} \\
& Atomic & 63.64\% & \colora{63.64\%} & \colora{90.91\%} & \colora{90.91\%} \\
& Minimal & \colorc{18.18\%} & \colora{81.82\%} & \colora{72.73\%} & \colora{54.55\%} \\
\hline
\multirow{5}{*}{Aggrega.} & Agreement rate & \colord{54.55\%} & 75.76\% & 81.82\% & 75.76\% \\
& Precision & 62.50\% & 80.95\% & 85.71\% & 74.07\% \\
& Recall & 71.43\% & 80.95\% & 85.71\% & 95.24\% \\
& Specificity & 25.00\% & 66.67\% & 75.00\% & 41.67\% \\
& F1 score & 66.67\% & \colore{80.95\%} & \colore{85.71\%} & \colore{83.33\%} \\
\hline
\end{tabular}
\end{table}

\subsection{ChatGPT-human agreement rate}

% OVERALL EVALUATION IS 75%
While performing the evaluation using ChatGPT, measures were taken to cover the rest of the metrics described in Dalpiaz et al.~\cite{aqusa}. In terms of agreement rate with human evaluators, ChatGPT's performance was relatively stable across the three assessments, as reported in Table~\ref{tab:threeruns}, with agreement rates ranging from \colord{73\% to 75\%}. However, this suggests that there is room for improvement in the agreement rate between ChatGPT and human evaluators. A 25\% error rate may be problematic in certain situations. As a result, enhancing ChatGPT's performance could increase its reliability and effectiveness in various applications.

\begin{table}[]
\centering
\caption{Agreement with human evaluations of ChatGPT, across 11 randomly sampled user stories, using different strategies to interpret the output}
\label{tab:threeruns}
\resizebox{\linewidth}{!}{
\begin{tabular}{llccccccc}
\hline
& & \multicolumn{3}{c}{ChatGPT} & \multicolumn{3}{c}{Interpretation strategy} \\
\textbf{Type} & \textbf{Metric} & \textbf{Run 1} & \textbf{Run 2} & \textbf{Run 3} & \textbf{AL1} & \textbf{BO3} & \textbf{PA3} \\
\hline
\multirow{7}{*}{Criteria} & Well-formed & \colora{81.82\%} & \colora{81.82\%} & \colora{81.82\%} & 81.82\% & 81.82\% & \textbf{88.89}\% \\
& Atomic & \colora{63.64\%} & \colora{90.91\%} & \colora{90.91\%} & 72.73\% & 90.91\% & \textbf{100.00}\% \\
& Minimal & \colora{81.82\%} & \colora{72.73\%} & \colora{54.55\%} & 45.45\% & 81.82\%  & \textbf{100.00}\% \\
& Conceptually sound & 90.91\% & 90.91\% & 63.64\% & 63.64\% & 81.82\%  & \textbf{100.00}\% \\
& Unambiguous & 54.55\% & 54.55\% & 63.64\% & \textbf{72.73}\% & 54.55\% & 62.50\% \\
& Full sentence & 45.45\% & 63.64\% & \textbf{81.82}\% & \textbf{81.82}\% & 63.64\% & 80.00\% \\
& Estimable & \textbf{90.91}\% & 72.73\% & 81.82\% & 72.73\% & 81.82\% & 88.89\% \\
\hline
\multirow{6}{*}{Aggrega.} & Agreement rate & \colord{72.73\%} & \colord{75.32\%} & \colord{74.03\%} & \colorc{70.13\%} & \colorc{76.62\%} & \colore{87.23\%} \\
& Precision & 83.33\% & 82.69\% & 77.05\% & 72.06\% & 81.82\% & \textbf{94.74}\% \\
& Recall & 75.47\% & 81.13\% & 88.68\% & \textbf{92.45}\% & 84.91\% & 90.00\% \\
& Specificity & 66.67\% & 62.50\% & 41.67\% & 20.83\% & 58.33\% & \textbf{71.43}\%\\
& F1 score & 79.21\% & 81.90\% & 82.46\% & 80.99\% & 83.33\% & \textbf{92.31}\%\\
& Coverage & 100.00\% & 100.00\% & 100.00\% & 100.00\% & 100.00\% & \colore{61.04\%} \\
\hline
\end{tabular}}
\end{table}

The agreement rates reported in Tables~\ref{tab:aqusa} and~\ref{tab:threeruns} include both true positives, where human raters and tools agreed on an overall positive evaluation of a user story, and true negatives, where humans and the tools agreed on a negative evaluation. Future work, though, might look into database entries where human raters and ChatGPT do not agree on the evaluations.

\subsection{How to select an answer based on ChatGPT's output}

% CHATGPT IS NOT VERY CONSISTENT, BUT WHEN CHATGPT IS CONSISTENT, CHATGPT IS VERY ACCURATE
In our study, we evaluated the consistency and reliability of ChatGPT in evaluating user stories against a set of predetermined criteria. Our results indicate that ChatGPT was consistent with itself in \colore{61\%} of the evaluations, meaning it gave the same response for a given pair \{criteria, user story\} in three separate runs (PA3). Furthermore, ChatGPT agreed with itself in at least two out of three runs in 83.11\% of the evaluations. These findings suggest that ChatGPT's evaluations are relatively stable when it comes to evaluating user stories. Moreover, we observed that in the subset of evaluations where ChatGPT was consistent across all three runs, the agreement rate with humans was \colore{87\%}, with precision and recall scores of 95\% and 90\%, respectively.

% THIS OPENS THE DOOR TO CONSIDER STRATEGIES ON HOW TO USE CHATGPT'S OUTPUT
The higher rate of agreement between humans and ChatGPT has encouraged us to explore stricter criteria for identifying positive responses. The use of a ``best of three'' approach in which ChatGPT is required to give a positive response in at least two out of three attempts resulted in a slightly higher agreement rate between humans and ChatGPT, reaching (\colorc{77\%}). However, when responses from ChatGPT were classified as positive if at least one positive response was given (AL1), there was more variability in ChatGPT's responses, leading to a decrease in the agreement rate to \colorc{70\%}.

\section{Discussion}

% ONLY 3 ROUNDS HERE BUT NEEDS FURTHER WORK
The agreement rate remained constant across the first three rounds, as evidenced by Tables~\ref{tab:aqusa} and~\ref{tab:threeruns}. Thus, we opted to conclude our testing after these three runs. However, to establish the reliability of these initial findings, additional evaluations of new user stories and more ChatGPT runs are necessary.

% THE CRITERIA IN WHICH CHATGPT STRUGGLE VARY A LOT ACROSS RUNS
Table~\ref{tab:threeruns} shows that the criteria with the highest and lowest agreement rates differed in the three runs, which suggests that certain criteria may have unclear definitions or abstract qualities that made it difficult for ChatGPT to consistently agree with human evaluators. To better understand this discrepancy, further research could examine the specific criteria that posed challenges for ChatGPT or where it exhibited inconsistencies.

% CONSISTENCY SEEMS TO BE IMPORTANT BUT COULD BE A BIAS
Although ChatGPT's consistency in generating responses may correlate with the level of agreement from human evaluators, it is important to note that consistency does not necessarily equate to the accuracy or appropriateness of the output. Additionally, the consistency could be due to potential bias present in the training data. To expand upon our findings, we aim to engage seasoned practitioners in the upcoming workshop and replicate the process we undertook in this pilot investigation. This study served as a preliminary exploration for the forthcoming workshop and we plan to gather additional data through the participation of these experienced professionals.

However, integrating ChatGPT into Agile development requires a thorough assessment of its capabilities, strengths, and limitations. While ChatGPT has shown promise in this task, its performance is not flawless, and it remains an emerging technology that is susceptible to potential biases and limitations. Therefore, careful consideration of ChatGPT's applicability and limitations is necessary before its integration into Agile software development processes~\cite{borji2023categorical}. On the other hand, GPT-4's expanded architectural model size might play a pivotal role in enhancing its proficiency in NLP, which could lead to increased accuracy and relevance in the generated responses~\cite{koubaa2023gpt}. 

However, a major obstacle to using ChatGPT in the requirements elicitation process is the issue of extrinsic hallucinations~\cite{bang2023multitask}. Non-experts who rely on AI systems might not possess the technical knowledge to evaluate the accuracy and reliability of the generated outputs in some cases. This issue highlights the importance of ensuring the trustworthiness of the AI systems being implemented. Ensuring trustworthiness in AI, particularly in the context of non-experts using ChatGPT for user story evaluation, requires careful consideration of several factors including transparency, explainability, bias mitigation, and continuous improvement through user feedback to be incorporated into the development and implemented process of these AI systems in such human-centric processes.

\section{Conclusion \& Future Work}

The study examines the effectiveness of ChatGPT in assessing user stories using predetermined boolean criteria, particularly when it produces consistent results across multiple evaluations. The research focuses on the agreement rate between humans and ChatGPT in evaluating user stories based on said criteria. The results indicate that ChatGPT is more capable of replicating human evaluation (approximately 75\%) than the AQUSA tool as demonstrated in T\~{o}emets~\cite{Temets2020AnalysingTQ}. 

While the model performs sufficiently well in independent runs, it exhibits inconsistency in its Boolean outputs when tested multiple times. This suggests caution in interpreting its evaluations and underlines the need for further research into the factors affecting ChatGPT's consistency and reliability.

To address the issue of unstable outputs, the paper suggests strategies such as selecting the ``best of three'' approach. However, the question of whether ChatGPT's raw outputs can be used directly by non-expert users raises important concerns about the trustworthiness of AI systems. As a result, high-level trustworthiness requirements must be established to ensure that ChatGPT and other AI tools are integrated into Agile software development processes following trustworthy AI principles. The integration of ChatGPT, into Agile software development processes, requires careful consideration of its limitations and strengths and the potential impact on the development process. Further research is needed to explore ways to ensure that ChatGPT and other AI systems can be used reliably and effectively in Agile development environments.

\bibliographystyle{ieeetr}
\bibliography{main}

\end{document}